\documentclass[aps,prd,superscriptaddress,showpacs,amsmath,amssymb]{revtex4}
\usepackage{graphicx,epsf}
\usepackage[]{latexsym}
\newcommand{\be}{\begin{eqnarray}}
\newcommand{\ee}{\end{eqnarray}}

\begin{document}
\title{Noncommutative Quantum Hall Effect and Aharonov-Bohm Effect}

\author{B. Harms}
\email{bharms@bama.ua.edu}
\affiliation{Department of Physics and Astronomy, The University
of Alabama, Box 870324, Tuscaloosa, AL 35487-0324, USA}
\author{O. Micu}
\email{micu001@bama.ua.edu}
\affiliation{Department of Physics and Astronomy, The University
of Alabama, Box 870324, Tuscaloosa, AL 35487-0324, USA}
\begin{abstract}
We study a system of electrons moving on a noncommutative plane in
the presence of an external magnetic field which is perpendicular
to this plane. For generality we assume that the coordinates and
the momenta are both noncommutative. We make a transformation from
the noncommutative coordinates to a set of commuting coordinates
and then we write the Hamiltonian for this system.  The energy
spectrum and the expectation value of the current can then be calculated
and the Hall conductivity can be extracted.
We use the same method to calculate the phase shift for the
Aharonov-Bohm effect. Precession measurements could allow
strong upper limits to be imposed on the noncommutativity coordinate and
momentum parameters $\Theta$ and $\Xi$.
\end{abstract}
\pacs{11.10.Nx, 73.43.-f}
\maketitle
\large
\section{Introduction}
\label{intro}

Noncommutative theories arise in string
theory~\cite{witten,seiberg} and in the present search for quantum
gravity~\cite{moffat}, while Yang-Mills theories on
non-commutative spaces~\cite{witten1} appear in string theory and
M-theory. The noncommutative theories which are studied the most
are the ones in which it is assumed that coordinates do not
commute with each other. For more generality we will assume that
the momenta are noncommutative as well. In the end if one wants to
restrict these results to the case where only the coordinates
are noncommutative, one can set the parameter that describes the
noncommutativity of the momenta to zero.
\par
We shall follow an approach \cite{smailagic} in which we will express the noncommutative
coordinates $x_i$ , $p_i$ as linear combinations of canonical
variables of quantum mechanics $\alpha_i$ , $\beta_i$. We will see
that the noncommutativity will introduce additional terms in the
Hamiltonian of the equivalent commutative description.
\par
In the present work we are calculating modifications to the
quantum Hall effect and to the Aharonov-Bohm effect in this
noncommutative scenario. In the former effect an electric current
flows through a conductor in a magnetic field which has a
component perpendicular to the plane of the electron's trajectory.
The magnetic field exerts a transverse force on the electrons
which tends to push them to one side of the conductor. This is
most evident in a flat and thin conductor where the magnetic field
is perpendicular to the plane of the conductor. Charge accumulates at the sides of
the conductors producing a measurable voltage between the two
sides of the conductor. The case of charged particles in magnetic
fields (the Landau problem) was previously considered in the
literature from prospectives which differ from the one taken in the present work \cite{nair},
\cite{duval}, \cite{kokado}, \cite{saha}, \cite{horvathy}.
\par
The Aharonov-Bohm effect emphasizes the fact that it is not the
electric and magnetic fields but the electromagnetic potentials
which are the fundamental quantities in quantum mechanics. In this
effect a beam of electrons is split in two and the two beams
follow two different paths.  An interference pattern is produced
when the two different beams of electrons recombine because there
will be a phase shift between the two beams, and this phase shift
depends on the magnetic flux enclosed by the two alternative
paths. This phase shift is observed even if they pass through
regions of space in which the magnetic field is null but the
vector potential is not zero. The noncommutative Aharonov-Bohm
effect was studied using the star product approach in
\cite{tureanu}, \cite{basu}.
\par
In Section II we consider electrons
which are moving on a noncommutative plane in the presence of an
electric field in this plane and an external magnetic field which
is perpendicular to the noncommutative plane. In the commutative
case the experiment described above leads to the Hall effect. Once
again, for more generality we assume both the coordinates (which
now become operators) and momenta do not commute. We calculate
corrections due to noncommutativity to the Hall conductivity, and
we will show that in the limit when the parameters describing
noncommutativity go to zero, we recover the commutative case. In
Section III we calculate deviations due to noncommutativity
to the phase shift for the Aharonov-Bohm effect. Also here we will
show that in the commutative limit we reproduce the usual
results.  Section IV contains the limits on the noncommutativity
parameters which we obtain from our analysis. In Section V we discuss the
results of our analysis.

\section{Noncommutative quantum hall effect}
\subsection{Hall effect}
\par An electron moving on the $(x,y)$ plane in
a uniform electric field $\vec{E}=-\vec{\nabla}\phi$ and a
uniform magnetic field $B$ which is perpendicular to the plane is
described by the Hamiltonian
\be
H=\frac{1}{2m}\left(\vec{p}+\frac{e}{c}\vec{A}\right)^2 -e\phi,
\label{hamiltonian1}
\ee
We will adopt the symmetric gauge (this
gauge is well suited for the experiment described in the previous
section)
\be
\vec{A}=\left(-\frac{B}{2}y,\frac{B}{2}x\right)
\label{gauge1}
\ee
and we will consider the scalar potential to be
\be
\phi=-E x \label{gauge2}
\ee
If we substitute (\ref{gauge1})
and (\ref{gauge2}) into (\ref{hamiltonian1}), we can write the
Hamiltonian in the following form
\be
H(\vec{p},\vec{r})=\frac{1}{2m}
\left[\left(p_x-\frac{eB}{2c}y\right)^2+\left(p_y+\frac{eB}{2c}x\right)^2\right]
+e E x,
\label{hamiltonian2}
\ee
\par
We want to calculate
modifications to the Hall conductivity due to the effects of space
noncommutativity. For generality, we assume a set of coordinates
and momenta which satisfy the following
commutation relations
\be
\left[x_i,x_j\right] &=& i\Theta_{ij},\label{commutator0}\\
\left[p_i,p_j\right] &=& i \hbar^2\Xi_{ij},\label{commutator1}\\
\left[x_i,p_j\right] &=&i\hbar\,\delta_{ij}. \label{commutator2}
\ee
where $\Theta_{ij}$ and $\Xi_{ij}$ are antisymmetric matrices
characterizing the noncommutativity of the phase space geometry.
\par Following the same treatment as \cite{smailagic} we
define linear transformations from
 the set of noncommutative coordinates to
 a commutative set of canonically conjugate coordinates $(\alpha_i,\beta_i)$ which satisfy
 \be
 \left[\alpha_i,\alpha_j\right] &=& 0,\label{commutators0}\\
\left[\beta_i,\beta_j\right] &=&0,\label{commutators1}\\
\left[\alpha_i,\beta_j\right] &=&i\hbar\,\delta_{ij}.
\label{commutators2}
\ee
\par
The relation between the two sets of
coordinates is defined as follows
\be
x_i&=&a_{ij}\alpha_j+b_{ij}\beta_j\label{transformations1}\\
p_i&=&c_{ij}\beta_j+d_{ij}\alpha_j\ \label{transformations2}
\ee
where a, b, c and d are in this case 2$\times$2 transformation
matrices. The relations (\ref{commutator0}), (\ref{commutator1}),
(\ref{commutator2}) and (\ref{commutators0}),
(\ref{commutators1}), (\ref{commutators2}) determine the
conditions which the transformation matrices must satisfy. In
matrix form they are
\be
\bf{a}\bf{b}^{T}-\bf{b}
\bf{a}^{T}&=& \frac{\bf{\Theta}}{\hbar}\label{matrix1}\\
\bf{c}\bf{d}^{T}-\bf{d}\bf{c}^{T}&=& -\hbar~\bf{\Xi}\label{matrix2}\\
\bf{c}\bf{a}^{T}-\bf{b}\bf{d}^{T}&=&\bf{I} \label{matrix3}
\ee
where \textbf{$\bf{\Theta}$} and \textbf{$\bf{\Xi}$} are
antisymmetric matrices.
\par
The transformation matrices are not unique (more details can be found in \cite{smailagic}) but a convenient
choice for our purposes during the calculations that follow is to
keep matrices \textbf{a} and \textbf{c} diagonal and single
valued. In order to maintain the same number of free
parameters, matrices \textbf{b} and \textbf{d} are chosen to be
antisymmetric
\be
a_{ij}\equiv a~\delta_{ij},~~~~c_{ij}\equiv
c~\delta_{ij}\\
b_{ij}\equiv b~\epsilon_{ij},~~~~d_{ij}\equiv d~\epsilon_{ij}
\ee
Equations (\ref{matrix1}-\ref{matrix3}) become
\be &a~b&=-\frac{\Theta}{2\hbar}\\
&c~d&=\frac{\hbar~\Xi}{2}\\&a~c&-b~d=1
\ee We solve for three
parameters
and we get
\be &b&=-\frac{\Theta}{2a\hbar}
\label{coef1}\\
&c&=\frac{1}{2a}\left(1\pm\sqrt{\kappa} \right), ~~~~\kappa\equiv 1-\Theta\Xi
\label{coef2}\\
&d&=\frac{\hbar a}{\Theta}\left(1\mp\sqrt{\kappa} \right)
\label{coef3}
\ee
\par Substituting (\ref{transformations1}), (\ref{transformations2}) and (\ref{coef1}), (\ref{coef2}),
(\ref{coef3}) into our Hamiltonian (\ref{hamiltonian2}), we can
rewrite it in the following way
\be
H(\vec{\alpha},\vec{\beta})=\frac{1}{2m}\left[h_1^2\left(\alpha_i\right)^2+h_2^2\left(\beta_i\right)^2-h_3\epsilon_{ij}\alpha_i\beta_j\right]+aeE\alpha_1-\frac{\Theta}{2a}eE\beta_2\label{hamiltonianNC1}\ee
with  \be
h_1^2&=&a^2\left[\frac{\hbar}{\Theta}\left(1\mp\sqrt{\kappa}\right)-\left(\frac{eB}{2c}\right)\right]^2 \label{h1}\\
h_2^2&=&\frac{\Theta^2}{4\hbar^2a^2}\left[\frac{\hbar}{\Theta}\left(1\pm\sqrt{\kappa}\right)-\left(\frac{eB}{2c}\right)\right]^2 \label{h2}\\
h_3&=&\left[\left(\frac{eB}{2c}\right)^2\Theta+\hbar^2~\Xi-\frac{\hbar
eB}{c}\right]\label{h3}
 \ee
 \par We make the following coordinate
 transformations
 \be
 &\beta_1&\rightarrow\beta_1\\
 &\beta_2&\rightarrow\beta_2-\frac{\frac{\Theta}{2a}eE}{2h_2^2}
 \ee
 and the Hamiltonian will take the following form
 \be
 H(\vec{\alpha},\vec{\beta})=\frac{1}{2m}\left[h_1^2\left(\alpha_i\right)^2+h_2^2\left(\beta_i\right)^2+h_3\epsilon_{ij}\alpha_i\beta_j\right]+h_4\alpha_1-h_5
 \ee
 where we have defined
 \be
 h_4&=&\pm 2eEa\frac{\sqrt{\kappa}}{\Theta\left[\frac{\hbar}{\Theta}\left(1\pm\sqrt{\kappa}\right)-\left(\frac{eB}{2c}\right)\right]}\\
 h_5&=&\frac{1}{2m}\frac{e^2E^2}{4\left[\frac{\hbar}{\Theta}\left(1\pm\sqrt{\kappa}\right)-\left(\frac{eB}{2c}\right)\right]^2}
\ee
 \par
 Now we have to discuss the eigenvalue problem
 \be
 \hat{H}\Psi=\mathcal{E}\Psi\label{schroedinger}
 \ee
 \par It is convenient to perform the change of variables
 \cite{dayi}
 \be
 \hat{z}&=&\alpha_1+i\alpha_2,\\
 \hat{p_z}&=&\frac{1}{2}(\beta_1-i\beta_2).
 \ee
 We define two sets of creation and annihilation operators
 \be
 b^{\dag}=-2ih_2\hat{p}_{\bar{z}}+h_1\hat{z}+\lambda,~~~~ b=2ih_2\hat{p}_{z}+h_1\hat{\bar{z}}+\lambda
 \ee
 and
 \be
 d^{\dag}=-2ih_2\hat{p}_{\bar{z}}-h_1\hat{z},~~~~ d=2ih_2\hat{p}_{z}-h_1\hat{\bar{z}}
 \ee
where $\lambda=\mp meE\sqrt{\kappa}/h_3$. These two sets of
operators commute with each other and satisfy the following
commutation relations
\be
\left[b,b^{\dag}\right]=2m\hbar\omega\\
\left[d^{\dag},d\right]=2m\hbar\omega
\ee
with $\omega=h_3/m$.
\par In terms of these operators the Hamiltonian is
\be
\hat{H}=
\frac{1}{4m}(bb^{\dag}+b^{\dag}b)-\frac{\lambda}{2m}(d^{\dag}+d)-\frac{\lambda^2}{2m}-h_5\label{hamiltonian3}
\ee
\par We observe that the Hamiltonian in (\ref{hamiltonian3}) is
composed of two mutually commuting parts
\be
\hat{H}=\hat{H}_{osc}-\hat{H_1}
\ee
\par
We will calculate the eigenvalues $\mathcal{E}$ and the
eigenfunctions $\Psi$ of the two commuting parts of the
Hamiltonian separately. For the harmonic oscillator part \be
\hat{H}_{osc}=\frac{1}{4m}(bb^{\dag}+b^{\dag}b) \ee the eigenvalue
value equation $\hat{H}_{osc}\Phi_n=\mathcal{E}_n^{osc}\Phi_n$ is
easily solved and it leads to a discrete spectrum \be
\Phi_n&=&\frac{1}{\sqrt{(2m\hbar\omega)^n n!}}(b^{\dag})^n|0>,\\
\mathcal{E}_n^{osc}&=&\frac{\hbar\omega}{2}(2n+1),~~n=0,1,2,...
\ee
\par The eigenvalue equation for $\hat{H_1}\phi_{\gamma}=\mathcal{E}_{\gamma}\phi_{\gamma}$ can be
analyzed in terms of eigenvalues of the operators $\alpha_i$ and
$\beta_i$ \be
\phi_{\gamma}&=&exp\left(-i(\gamma\alpha_2+\frac{h_1}{\hbar
h_2}\alpha_1 \alpha_2)\right)\\
\mathcal{E}_{\gamma}&=&\frac{\hbar\lambda
h_2}{m}\gamma+\frac{\lambda^2}{2m}+h_5,~~ \gamma\in \mathbb{R} \ee
which is a continuous spectrum.
\par We can now combine the two solutions and as a result, the eigenfunctions and energy spectrum of the Hamiltonian
$\hat{H}$ are
\be
\Psi_{(n,\gamma,\Theta,\Xi)}=|n,\gamma,\Theta,\Xi\rangle&=&\frac{1}{\sqrt{(2m\hbar\omega)^n
n!}}exp\left(-i(\gamma\alpha_2+\frac{h_1}{\hbar
h_2}\alpha_1 \alpha_2)\right)(b^{\dag})^n|0>,\\
\mathcal{E}_{(n,\gamma)}&=&\frac{\hbar\omega}{2}(2n+1)-\frac{\hbar\lambda
h_2}{m}\gamma-\frac{\lambda^2}{2m}-h_5
\ee
\subsection{Hall conductivity}
The Hall conductivity can be calculated by means of the Hamiltonian
$\hat{H}$ given above. We define the current operator $\hat{\vec{J}}$ on the
noncommutative plane as
\be
\hat{\vec{J}}=\frac{ie\rho}{\hbar}\left[\hat{H},\hat{\vec{r}}\right]
\ee
\par The expectation values of the components of the current operator
$\langle\hat{\vec{J}}\rangle$ calculated with respect to the
eigenstates $|n,\gamma,\Theta,\Xi\rangle$ are
\be
\langle\hat{J_x}\rangle=0,\\
\langle\hat{J_y}\rangle=-e\rho\frac{1-\Theta\Xi}{\frac{B}{c}-\frac{eB^2\Theta}{4\hbar
c^2}-\frac{\hbar~\Xi}{e}}E.
\ee
\par
Therefore the Hall conductivity on the noncommutative plane,
which we denote by $\sigma_H$ is
\be
\sigma_H=-e\rho\frac{1-\Theta\Xi}{\frac{B}{c}-\frac{eB^2\Theta}{4\hbar
c^2}-\frac{\hbar~\Xi}{e}}
\ee
\par
We notice that if the noncommutativity parameters
$\Theta$ and $\Xi$ are taken to be equal to zero, we obtain the
same value for the Hall conductivity
 as in the commutative case ($\sigma_H=-\rho ec/B$)
\par
Using Eq.(52) we can look at particular cases derived from our
result. In the noncommutative scenarios which are most commonly
discussed in the literature, only the coordinates are
noncommutative and the parameter $\Xi$ is equal to zero.  In this
case we obtain \be
\sigma_H=-e\rho\frac{1}{\frac{B}{c}-\frac{eB^2\Theta}{4\hbar c^2}}
\label{theta1}\ee Alternatively we can imagine a noncommutative
scenario in which coordinates commute but momenta are
noncommutative. In this case also we obtain modifications of the
commutative Hall effect due to the presence of the term
proportional to $\Xi$
\be
\sigma_H=-e\rho\frac{1}{\frac{B}{c}-\frac{\hbar~\Xi}{e}}
\ee
\par From an experimental point of view, the last case that we
consider might be much more important. Our
result predicts that if space and momenta are noncommutative, even
without the presence of an external magnetic field, the value of
the Hall conductivity should be different from zero
\be
\sigma_H=e^2\rho\frac{1-\Theta\Xi}{\hbar~\Xi} \label{sigma}
\ee
In
this case the sign of the conductivity is different. If the
sensitivity of the experiments is increased sufficiently, the
effect might eventually be detected and it would be a clear
signature of noncommutativity. There is one situation in which the
Hall conductivity is still zero in this scenario, and that is if
the two noncommutativity parameters $\Theta$ and $\Xi$ are
naturally adjusted such that one is the inverse of the other. In
this case the numerator of (\ref{sigma}) would be equal to zero.
\section{Noncommutative Aharonov-Bohm effect}
We will use the same approach to study the Aharonov-Bohm effect in
the noncommutative plane. We start from a similar Hamiltonian as
in (\ref{hamiltonian1})
\be
H=\frac{1}{2m}\left(\vec{p}+\frac{e}{c}\vec{A}\right)^2+V
\label{hamiltonian4}
\ee
where $e$ is the charge on an electron.
\par Using the same gauge as in
(\ref{gauge1}), the Hamiltonian becomes
\be
H(\vec{p},\vec{r})=\frac{1}{2m}
\left[\left(p_x-\frac{eB}{2c}y\right)^2+\left(p_y+\frac{eB}{2c}x\right)^2\right]
+V(\vec{x}), \label{hamiltonian3}
\ee
\par
Following a similar derivation as in the previous part of the
paper, we can rewrite $H$ once again in terms of $\alpha_i$ and
$\beta_i$ as
\be
H(\vec{\alpha},\vec{\beta})=\frac{1}{2m}\left[h_1^2\left(\alpha_i\right)^2
+h_2^2\left(\beta_i\right)^2-h_3\epsilon_{ij}\alpha_i\beta_j\right]+V(\vec{\alpha},\vec{\beta})\label{hamiltonianAB1}
\ee
where the coefficients $h_1$, $h_2$ and $h_3$ are the ones defined
in  (\ref{h1}), (\ref{h2}) and (\ref{h3}). We use the fact that
$2h_1 h_2=h_3$ and we rewrite the Hamiltonian once again as
\be
H(\vec{\alpha},\vec{\beta})=\frac{1}{2m}\left[\left(h_2\vec{\beta}-h_1\vec{\alpha}\right)^2\right]
+V(\vec{\alpha},\vec{\beta})\label{hamiltonianAB2}
\ee
where we have defined $\vec{\beta}=(\beta_1,\beta_2)$ and
$\vec{\alpha}=(-\alpha_2,\alpha_1)$.
\par We now have to solve Schrodinger's equation
 \be
 \left[\frac{1}{2m}\left( h_2\frac{\hbar}{i}\nabla-h_1\vec{\alpha}\right)^2+V(\vec{\alpha},\vec{\beta})\right]\Psi=i\frac{\partial\Psi}{\partial
t}\label{schrodingerAB}
\ee
where we substituted
$\vec{\beta}\rightarrow\frac{\hbar}{i}\nabla$.
\par
We write
\be
\Psi=e^{ig}\psi \label{psi}
\ee
where
\be
g(\vec{\alpha})=\frac{h_1}{h_2 \hbar}\oint\vec{\alpha'}d\alpha'
\ee
\par
Then
\be
\nabla\Psi=e^{ig}(i\nabla g)\psi+e^{ig}\nabla\psi
\ee but
\be
\nabla g=\frac{h_1}{h_2 \hbar}\vec{\alpha}
\ee
and using this we have from Eq.(63)
\be
\left(
h_2\frac{\hbar}{i}\nabla-h_1\vec{\alpha}\right)\Psi=h_2\frac{\hbar}{i}e^{ig}\nabla\psi
\ee
The first term in the (\ref{schrodingerAB}) becomes
\be
\left(
h_2\frac{\hbar}{i}\nabla-h_1\vec{\alpha}\right)^2\Psi=-h_2^2\hbar^2
e^{ig}\nabla^2\psi\label{psi1}
\ee
\par
If we substitute (\ref{psi}) and (\ref{psi1}) into (\ref{schrodingerAB}) we can see that $g(\vec{\alpha})$ is just a phase difference and it will be equal to
\be
g(\vec{\alpha})&=&\frac{h_1}{h_2
\hbar}\oint^{\alpha}\vec{\alpha'}d\alpha'\nonumber\\
&=&\frac{h_1}{h_2\hbar}\oint^{\alpha}(-{\bf{i}}\alpha'_2+{\bf{j}}\alpha'_1)d\alpha'\nonumber\\
&=&\frac{h_1}{h_2\hbar} 2\pi\alpha^2
\ee
Substituting the expressions for the
constants $h_1$ and $h_2$ from (\ref{h1}) and (\ref{h2}) we have
for the phase shift
\be
g(\vec{\alpha})=\frac{2a^2\left[\frac{\hbar}{\Theta}\left(1\mp\sqrt{\kappa}\right)-\left(\frac{eB}{2c}\right)\right]}
{\Theta\left[\frac{\hbar}{\Theta}\left(1\pm\sqrt{\kappa}\right)-\left(\frac{eB}{2c}\right)\right]}
2\pi\alpha^2
\ee
\par
We notice that if we take the
noncommutativity parameters $\Theta$ and $\Xi$ to zero, only the
solution with the upper signs is physical
\be
g(\vec{\alpha})=\frac{2a^2\left[\frac{\hbar}{\Theta}\left(1-\sqrt{\kappa}\right)-\left(\frac{eB}{2c}\right)\right]}
{\Theta\left[\frac{\hbar}{\Theta}\left(1+\sqrt{\kappa}\right)-\left(\frac{eB}{2c}\right)\right]}
2\pi\alpha^2
\label{phase}
\ee
 Moreover for $\Theta$ and $\Xi$ zero,
setting  the free parameter $a$ to 1, gives the same phase
shift as in the commutative case $-e\Phi/\hbar$ where by $\Phi$
is the magnetic flux.
\par
In the limit when $\Theta$ and $\Xi$ are small, we can expand
$g(\vec{\alpha)}$ to the first order in both parameters, \be
g(\vec{\alpha})=
\left(-\frac{eB}{c\hbar}+\Xi-\frac{eB^2\Theta}{4c^2\hbar^2}\right)\pi\alpha^2
\label{phaseshift}\ee
\par
Again if we consider the case for which only the
coordinates are noncommutative but the momenta commute with each
other ($\Xi=0$), the phase shift becomes
\be
g(\vec{\alpha})=
\left(-\frac{eB}{c\hbar}-\frac{eB^2\Theta}{4c^2\hbar^2}\right)\pi\alpha^2
\ee
 Another scenario we can consider is the one for which
the coordinates commute but the momenta are noncommutative. In this
case  we again obtain modifications of the commutative phase shift
due to the presence of the term proportional to $\Xi$
\be
g(\vec{\alpha})= \left(-\frac{eB}{c\hbar}+\Xi\right)\pi\alpha^2
\ee
For these two last cases that we studied, we observe that if
the external magnetic field can be adjusted with enough
sensitivity, we should find a value of the magnetic field for
which the phase shift vanishes.
\par
In the absence of an external magnetic field, again the phase shift is different from zero if
the coordinates and the momenta are noncommutative. To see this we
look back at equation (\ref{phase}) and we set the magnetic field
to zero
 \be
g(\vec{\alpha})=\frac{2a^2\left(1-\sqrt{\kappa}\right)}
{\Theta\left(1+\sqrt{\kappa}\right)} 2\pi\alpha^2\label{phase1}
\ee

\section{Experimental limits on $\Theta$ and $\Xi$}

The Hall conductivity (\ref{theta1}) can be measured with an
accuracy of one part in a billion.  We can use this experimental limit to
impose an upper limit of $10^{-34}m^2$ on
the noncommutativity parameter $\Theta$. This limit on $\Theta$ is weaker by six orders of
magnitude than the one imposed by \cite{carroll} using data from
experiments which test Lorentz invariance. Also the authors of
\cite{falomir} propose a stronger limit on theta by measuring
differential cross sections for small angles in scattering
experiments. However, the latter experiment is very
difficult to perform because it requires the measurement of
scattering angles between 1 and 2 degrees at energies of the order
of ~200 GeV. We can also impose a limit on $\Theta$ using
Aharonov-Bohm measurements, but this limit is much weaker than the
one that can be imposed using the Hall effect.
\par
One of the advantages of our calculation is that we are able to consider
the case when momenta are noncommutative, and we can also impose
limits on the magnitude of the parameter which describes it.
>From the experiments which measure the Hall conductivity we find
that $\Xi$ must be smaller than $10^{-19}m^{-2}$. Also from
(\ref{phaseshift}) we can see that noncommutativity of momenta
would induce a phase shift even in the absence of external
magnetic fields.  If the momenta are noncommutative and the phase
shift could be measured with enough accuracy,   a
phase shift would be detected even in the absence of a magnetic field.

\section{Discussion}
The electrons in the low-temperature Hall effect and in the Aharonov-Bohm effect do not
interact appreciably with other particles, thus allowing  a rather simple form for the
Hamiltonian in each case.  Although noncommutativity introduces some complexity into the
expressions for the Hamiltonians in these two effects, they are still sufficiently simple
that we were able to obtain wave vectors and energy eigenvalues or the phase shifts in closed form.
This in turn allowed us to obtain  expressions for the quantities measured in each of
these two effects plus the deviations from the commutative forms of the quantities due to
noncommutativity in analytical form.   Several interesting features arise in the noncommutative
forms of the quantum Hall effect and the Aharonov-Bohm effect.  In the former effect the
deviation of the conductivity due to noncommutativity is independent of the magnetic field to lowest
order in the parameter $\Theta$ for $\Xi = 0$.  If the conductivity can be measured with sufficient
precision, a deviation from the normal (magnetic field dependent) value would be circumstantial
evidence for noncommutativity.  In both the quantum Hall effect and the Aharonov-Bohm effect deviations
occur in the conductivity and phase shift respectively even if there is no magnetic field.
\par
The limits which the two effects studied in this work can set on the $\Theta$ noncommutative parameter,
while not as strong as the one set in \cite{carroll}, are nevertheless significant.  The limit on
the $\Xi$ parameter which was obtained in this work is the first one we have seen. The results reported
on here suggest that high precision measurements in atomic and molecular systems may be able to rival or
even exceed the strongest limit set by nuclear systems \cite{carroll}. For this reason we are in the process of studying
the Josephson effect using the method described above.

\section{Acknowledgements}
This research was supported in part by the University of Alabama's Research Advisory Committee.
\end{document}